\documentclass[sigconf]{acmart}
\usepackage{multirow}
\usepackage[table,xcdraw]{xcolor}
\usepackage{graphicx}
\usepackage{booktabs}
\AtBeginDocument{%
  }

\setcopyright{acmlicensed}
\copyrightyear{2018}
\acmYear{2018}
\acmDOI{XXXXXXX.XXXXXXX}
\acmConference[Conference acronym 'XX]{Make sure to enter the correct
  conference title from your rights confirmation email}{June 03--05,
  2018}{Woodstock, NY}
\acmISBN{978-1-4503-XXXX-X/2018/06}

\begin{document}

\title{From Head to Tail: Asymmetric Knowledge Transfer in Long-tail Recommendation with Generative Semantic IDs}



\author{
  Chenyi Yan\textsuperscript{1}, 
  Ruocong Tang\textsuperscript{1}, 
  Xing Fang\textsuperscript{1}, 
  Yang Huang\textsuperscript{1}, 
  He Guo\textsuperscript{2}, 
  Jing Wang\textsuperscript{1}
}

\affiliation{%
  \institution{
    \textsuperscript{1}Alibaba Group, Hangzhou, China\\
    \textsuperscript{2}Beijing University, Beijing, China\\
    \{yanchenyi.ycy, tangruocong.trc, fangxing.fx, hy234680, jing.wangj1\}@taobao.com, \\
    2401210268@stu.pku.edu.cn
  }
  \country{} 
}







\renewcommand{\shortauthors}{Yan et al.}

\begin{abstract}
Long-tail recommendation in real-world e-commerce platforms remains challenging due to severe data imbalance. Existing methods often struggle to combine content-based multimodal features with collaborative signals. Many of these methods also ignore an important asymmetry in knowledge transfer between head and tail IDs: noisy signals from tail IDs can hurt representation learning for head IDs. This paper presents AKT-Rec, a framework for \textbf{A}symmetric \textbf{K}nowledge \textbf{T}ransfer in long-tail \textbf{Rec}ommendation that uses LLM-generated semantic IDs. AKT-Rec uses Multimodal LLMs (MLLMs) with supervised fine-tuning to align content representations with collaborative information for both items and users, producing semantic representations. It then discretizes these representations into semantic IDs with a Residual-Quantized VAE (RQ-VAE), which yields semantic clusters of similar entities. AKT-Rec has two main components: (1) \textbf{Cluster-Guided Adaptive Embedding}, which decomposes each ID representation into a cluster-level embedding that captures shared semantics and an individual embedding. Through an asymmetric contrastive objective and an activity-aware gating mechanism, this module directs knowledge transfer from head to tail IDs. (2) \textbf{Hierarchical Feature Aggregation}, which builds parallel feature views and adaptively fuses them to optimize predictions for samples with varying activity levels. Extensive experiments on a large-scale industrial dataset and online A/B testing on the Alibaba Tmall platform demonstrate the effectiveness of AKT-Rec. AKT-Rec improves offline performance by \textbf{0.35\%} in AUC and \textbf{1.53\%} in GAUC, outperforming several competitive baselines. In online A/B testing, AKT-Rec achieves a \textbf{2.76\%} increase in CTR and a \textbf{3.47\%} increase in GMV, validating its utility in real-world production environments.
\end{abstract}

\ccsdesc[500]{Information systems~Content ranking}

\keywords{Recommendation System, Multi-Modal, Large Language Model, Semantic ID}


\maketitle

\section{Introduction}
With the rapid expansion of e-commerce platforms, the scale of candidate item pools and user populations in recommender systems is growing at an unprecedented rate. However, this growth has also intensified the long-tail issue caused by imbalanced data distributions: a small fraction of popular (head) items receive most exposures, and a relatively small group of highly active users accounts for a large share of observed interests~\cite{dai2021poso,jangid2025deep,li2017two}. Such skewed distributions can severely impairs the performance of recommender systems, making it difficult to learn reliable representations for the many tail items and to capture the diverse preferences of tail users.
Although long-tail recommendation has been extensively studied over the past decade, existing methods still exhibit notable limitations. 
Graph-based models represent collaborative signals with graph structures (e.g., user-item interaction graphs) and use Graph Convolutional Networks (GCNs) to propagate information to tail nodes~\cite{luo2023improving,volkovs2017dropoutnet,cui2019class,zhao2023long,wei2023meta}. However, these methods are constrained by the inherent data sparsity of long-tail items in real-world scenarios and rely on manually specified graph construction rules, which often introduce redundant or noisy edges. 
Sample augmentation models generate synthetic pseudo-interactions for tail items to narrow the distribution gap between tail and head items~\cite{qin2024gacrec,liu2020deep}. Nevertheless, these synthetic samples inevitably can distort the underlying data distribution, and their quality is hard to control, often leading to suboptimal performance.
More recently, motivated by progress in Large Language Models (LLMs), several studies use LLMs to extract rich content features for items to replace or augment collaborative signals, thereby facilitating the modeling of long-tail items\cite{jia2025learn,sheng2024language,baek2024knowledge,wei2024llmrec,du2024enhancing}. In practice, these content features are often underutilized in models still dominated by collaborative signals, leading to limited gains. Moreover, many of these approaches focus on tail IDs in isolation while overlooking the potential negative impact on head items. Head items can be accurately modeled using abundant interaction data, and this rich data can be leveraged to support the representation learning of similar long-tail IDs.

\begin{figure*}[t]
    \centering
    \includegraphics[width=0.93\textwidth]{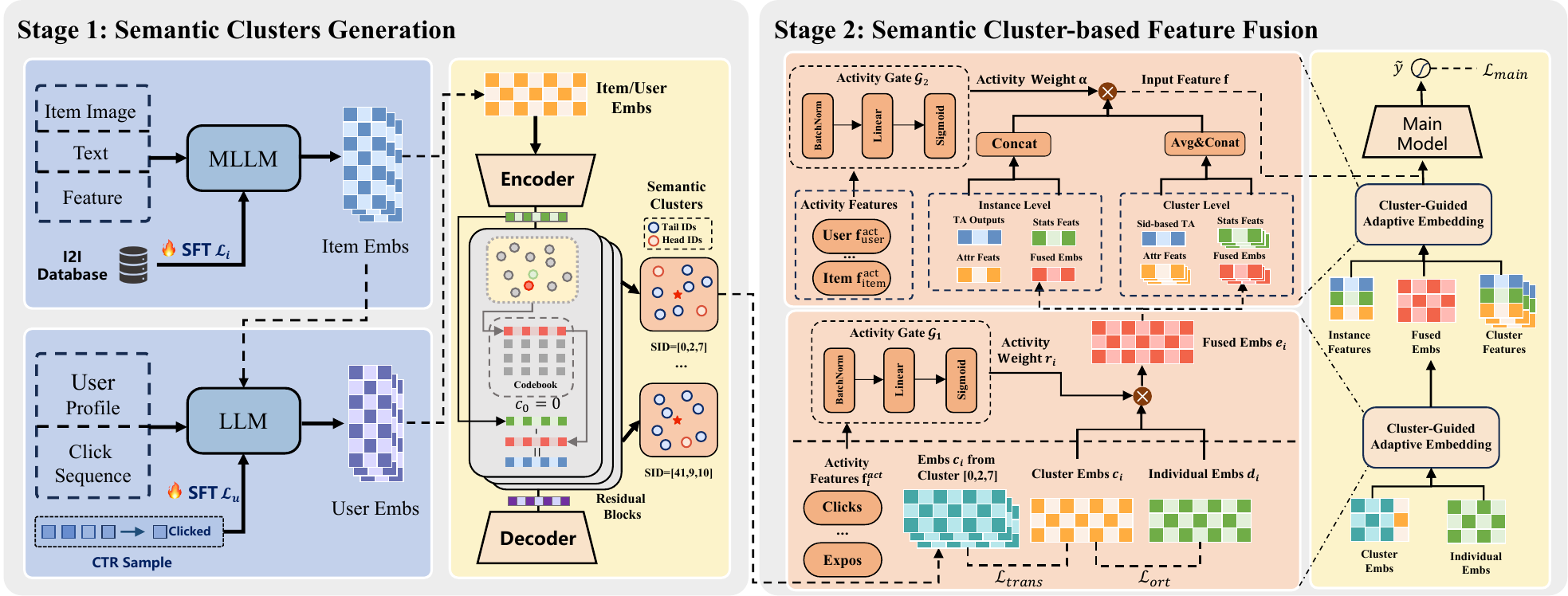}
    \caption{The Overall architecture of AKT-Rec}
    \label{fig:framework}
\end{figure*}

In this paper, to address the aforementioned challenges, we propose \textbf{AKT-Rec} a long‑tail recommendation framework based on LLM‑generated semantic IDs. It integrates multi-modal and collaborative features and enables asymmetric knowledge transfer from head to tail users/items. As illustrated in Figure~\ref{fig:framework}, the recommendation pipeline is organized into two stages.
In the first stage, AKT-Rec employs a Multimodal Large Language Model (MLLM) to extract item and user representations based on item-to-item co-occurrence relationships. A Residual-Quantized Variational Autoencoder (RQ-VAE) then clusters items and users separately to produce their corresponding semantic IDs. The quantization procedure is configured to yield a high collision rate such that each semantic ID is shared by multiple similar items or users. 
In the second stage, each item or user is represented by two embeddings: a \textit{cluster embedding} that encodes the shared semantics within a cluster and an \textit{individual embedding} that captures ID-specific information. We introduce an activity-aware asymmetric InfoNCE objective to transfer knowledge from head IDs to tail IDs. Moreover, we design a loss function to encourage disentanglement between the two embeddings and to reduce redundant information. Finally, we design a feature aggregation module for the semantic clusters and employ an additional fusion network to integrate these cluster-level features with traditional item/user features.

Our contributions are summarized as follows:
\begin{itemize}
    \item We propose \textbf{AKT-Rec}, a novel LLM-based framework for long-tail recommendation. It leverages semantic clusters to transfer knowledge from head IDs to tail IDs while ensuring that the representation learning of head items is not adversely affected by tail items, thereby mitigating the long-tail problem in recommender systems.
    
    \item We introduce an adaptive, activity-aware embedding mechanism that assigns each ID both a cluster representation and an individual representation, and dynamically adjusts the balance between them. Furthermore, we develop a semantic cluster-aware sequential aggregation module that enables efficient and accurate information aggregation.
    
    \item We conduct extensive experiments on a large-scale industrial dataset. The proposed \textbf{AKT-Rec} achieves a significant improvement in online A/B testing in our production environment, demonstrating its effectiveness and reliability.
\end{itemize}


\section{Problem Formulation}

We formulate the long-tail recommendation task as a Click-Through Rate (CTR) prediction problem over user--item interaction data exhibiting long-tail distributions, including both long-tail users and long-tail items. Let $\mathcal{U}$ denote the set of users and $\mathcal{I}$ denote the set of items. For any user $u \in \mathcal{U}$ and item $i \in \mathcal{I}$, we define the interaction label
\[
y_{u,i} = 
\begin{cases}
1, & \text{if user } u \text{ clicks item } i, \\
0, & \text{otherwise}.
\end{cases}
\]

We assume that both user activity levels and item popularity follow long-tail distributions. Specifically, the interaction frequency of user $u$ and item $i$ are defined as
\[
f(u) = \sum_{i \in \mathcal{I}} y_{u,i}, \quad
i = \sum_{u \in \mathcal{U}} y_{u,i},
\]
and both $f(u)$ and $i$ exhibit a heavy-tailed distribution.

Our goal is to leverage the rich interaction histories from head users and items to improve CTR prediction accuracy on long-tail users and items, while preserving the prediction accuracy for head users and items.

\section{Methodology}
\subsection{Semantic Clusters Generation}
\subsubsection{Representation Extraction}
To generate high-quality representations, we design a two-stage extraction paradigm that first produces item embeddings and subsequently derives user embeddings by integrating historical behaviors with attributes. We employ a pre-trained MLLM (e.g. ~\cite{zhang2024gme}) to extract item representations. The prompt includes the item image, the textual description, and statistical features such as $N$-day click-through rates (CTR), conversion rates (CVR), and add-to-cart rates. These features are quantized into discrete levels for model understanding. To incorporate collaborative signals, we identify frequently co-occurring item pairs $\langle i_1, i_2 \rangle$ and utilize contrastive learning to align their representations via the InfoNCE loss:
\begin{equation}
\mathcal{L}_{i} = -\frac{1}{N} \sum_{i=1}^{N} \log \frac{\exp(\mathbf{z}_i\cdot \mathbf{z}_i^+ / \tau)}{\exp(\mathbf{z}_i \cdot \mathbf{z}_i^+ / \tau) + \sum_{j=1}^{K} \exp(\mathbf{z}_i \cdot \mathbf{z}_j^- / \tau)}
\end{equation}
where $N$ denotes the batch size, $K$ represents the number of negative samples, $\mathbf{z}_i$ is the embedding of the $i$-th sample, $\mathbf{z}_i^+$ is the corresponding positive embedding, $\mathbf{z}_j^-$ are negative embeddings, and $\tau$ is the temperature hyperparameter.

Subsequently, another LLM generates user semantic representations from interaction history and profiles, following the supervised approach in ~\cite{jiang2025large}. The prompt incorporates user attributes and chronological interaction sequences (e.g., clicks within 30 days) with categorical and statistical data. The model generates an interest token to capture the user’s future preferences and predicts the corresponding item category. We fine-tune the LLM using user profiles, historical click sequences, and ground-truth items from traditional CTR task samples. The objective $\mathcal{L}_u$ is formulated as:
\begin{equation}
\mathcal{L}_u = -\text{sim}(\hat{\mathbf{e}}_{t}, \mathbf{e}_t)-
\left[ y_c \log(\hat{y}_{c}) + (1 - y_c) \log(1 - \hat{y}_{c}) \right]
\end{equation}
where $\hat{\mathbf{e}}_{t}$ is the hidden vector of the interest token serving as the user semantic representation, $\hat{y}_{c}$ is the predicted category, $\mathbf{e}_t$ and $y_c$ represent the clicked item's semantic representation and its category, and $\text{sim}(\cdot)$ denotes cosine similarity.

\subsubsection{Cluster Generation}
We use a RQ-VAE to quantize semantic representations into discrete identifiers, forming a coarse-to-fine hierarchical structure. Given an encoder $E$, a decoder $D$, and a residual $r_0$ initialized as input $x$, the quantization process involves $N$ codebook layers of size $M$ with vectors $\{e_i\}_{i=1}^M$. Identifiers are derived iteratively:
\begin{equation}
id_k = \text{argmin}_i ||r_{k-1} - e_i||^2, \quad r_k = r_{k-1} - e_{id_k}, \quad 1 \leq k \leq N
\end{equation}
For any representation, the $N$-layer RQ-VAE generates a sequence $(id_1, id_2, \dots, id_N)$ via nearest neighbor search, which forms the final hierarchical semantic ID. We calibrate the layers and codebook size so a single ID can represent multiple similar items, facilitating semantic cluster formation.

\subsection{Semantic Cluster-based Feature Fusion}
We design a decoupled module at both the embedding and feature levels to preserve information at different granularities and facilitate asymmetric knowledge transfer from head to tail items.

\subsubsection{Cluster-Guided Adaptive Embedding}
We introduce a semantic cluster-guided adaptive embedding mechanism that decomposes the representation of each ID into two components: a cluster embedding $\mathbf{c}_{i} \in \mathbf{R}^m$ and an individual embedding $\mathbf{d}_i \in \mathbf{R}^m$. We align cluster embeddings of IDs within the same semantic cluster with contrastive learning. To control the knowledge transfer between head and tail IDs, we use an asymmetric InfoNCE objective
\begin{equation}
\mathcal{L}_{\text{trans}} = \lambda_1 \mathcal{L}_{\text{info}}(\mathbf{c}_{i}^{\text{head}}, sg(\mathbf{c}_{i}^{\text{tail}})) + \lambda_2 \mathcal{L}_{\text{info}}(\mathbf{c}_{i}^{\text{tail}}, sg(\mathbf{c}_{i}^{\text{head}}))
\end{equation}
where $\mathcal{L}_{\text{info}}(x, y)$ denotes the InfoNCE loss using $y$ as the positive sample, $\mathbf{c}_{i}^{\text{head}}$ and $\mathbf{c}_{i}^{\text{tail}}$ represent the cluster embeddings of head and tail IDs respectively, and $sg(\cdot)$ denotes the stop-gradient operation. Normally, we set $\lambda_1 < \lambda_2$ to ensure that knowledge predominantly transfers from head to tail items. To avoid information redundancy between $\mathbf{c}_{i}$ and $\mathbf{d}_i$ which could lead to optimization collapse, we introduce a soft orthogonality regularizer to encourage the encoding of complementary information:
\begin{equation}
\mathcal{L}_{\text{ortho}} = \left( \frac{\mathbf{c}_{i}^\top \mathbf{d}_i}{\|\mathbf{c}_{i}\|_2 \cdot \|\mathbf{d}_i\|_2} \right)^2
\end{equation}
The final representation for a user or item is an adaptive fusion of these two embeddings based on activity features
\begin{equation}
    r_i  = \mathcal{G}_1(\mathbf{f}_i^{\text{act}}),\ \mathbf{e}_i  = r_i \cdot \mathbf{c}_{i} + (1 - r_i) \cdot \mathbf{d}_i
\end{equation} 
where $\mathbf{f}_i^{\text{act}}$ represent the activity features of the ID and $\mathcal{G}_1$ is a  feedforward neural network and $\mathbf{e}_i$ denotes the fused embedding.

\subsubsection{Hierarchical Feature Aggregation}
To exploit the hierarchical structure of semantic clusters, we construct two parallel views at different feature levels: the instance level and the cluster level.

The instance level focuses on the fine-grained context of specific users and items during an interaction. Features at this level include the individual embeddings $\mathbf{e}_i$ and $\mathbf{u}_i$, along with attributes and statistical features represented as embeddings. We define the user and item feature vectors as follows \begin{equation}
    \mathbf{H}_u =[\mathbf{u}_{\text{attr}};\mathbf{u}_{\text{stats}};\mathbf{u}_i ],\ 
    \mathbf{H}_i =[\mathbf{i}_{\text{attr}}, \mathbf{i}_{\text{stats}}, \mathbf{e}_i]
\end{equation}
The user interaction history $\mathbf{s}_{u} = [\mathbf{h}_0, \mathbf{h}_1,\cdots,\mathbf{h}_L]$ is encoded with the candidate item $\mathbf{e}_i$ using target-aware attention~\cite{zhou2018deep}
\begin{equation}
    \mathbf{S}_{u,i} = \sum_{j=1}^{L} \alpha_{i,j} \mathbf{h}_j,\  
    \alpha_{i,j} = \frac{\exp(\mathbf{e}_i \mathbf{h}_j^T)}{\sum_{k=1}^{L} \exp(\mathbf{e}_i \mathbf{h}_k^T)},
\end{equation}
These components are concatenated to form the instance-level representation $\mathbf{H}_{\text{inst}}$, ensuring prediction precision through maximum granularity.

The cluster level captures the representative context of the user cluster $G(u)$ and the item cluster $G(i)$. We compute cluster-level features by averaging the instance features within each cluster
\begin{equation}
    \mathbf{H}_{G(u)} = \frac{\sum_{{u'} \in G(u)}(\mathbf{H}_{u'})}{\|{G(u)}\|},\ 
    \mathbf{H}_{G(i)} = \frac{\sum_{{i'} \in G(i)}(\mathbf{H}_{i'})}{\|{G(i)}\|}
\end{equation}
aggregating behaviors from all users in a cluster leads to sequences that are too long for the latency constraints of online serving. Following~\cite{pi2020search}, we therefore use a hard retrieval strategy based on the top-level semantic ID to fetch behaviors that are most relevant to the candidate item. A target attention mechanism is then applied to this cluster-level sequence to generate $\mathbf{S}_{G(u),i}$. These features are concatenated into the cluster-level representation $\mathbf{H}_{\text{clust}}$.

\subsubsection{Adaptive Feature Fusion}
Rather than simple concatenation, we employ a gating network based on joint user-item activity to adaptively balance the contributions of the two hierarchical views. Given user activity features $\mathbf{f}_{user}^{\text{act}}$, item activity features $\mathbf{f}_{item}^{\text{act}}$, and cross-features $\mathbf{f}_{cross}^{\text{act}}$, the fusion weight is calculated as 
\begin{equation}
    \alpha = \mathcal{G}_2([\mathbf{f}_{user}^{\text{act}};\mathbf{f}_{item}^{\text{act}};\mathbf{f}_{cross}^{\text{act}}])
\end{equation}
where $\mathcal{G}_2$ is a  feedforward neural network. The final feature input is formulated as $\mathbf{f} = \alpha \cdot \mathbf{H}_{\text{clust}} + (1 - \alpha) \cdot \mathbf{H}_{\text{inst}}$. This combined feature $\mathbf{f}$ is then fed into the ranking network $\mathcal{F}$, such as a multilayer perceptron network or a multi-gate mixture-of-experts model~\cite{ma2018modeling} to predict click preference.
\begin{equation}
    \hat{y}=\mathcal{F}(\mathbf{f}),\ \mathcal{L}_{\text{main}}=- \left[ y \log(\hat{y}) + (1 - y) \log(1 - \hat{y}) \right]
\end{equation}
where $y$ is the click label. The CTR model is trained by minimizing the final loss 
\begin{equation}
    \mathcal{L}_{\mathrm{ctr}}=\mathcal{L}_{\text{main}}+\mathcal{L}_{\text{trans}}+\lambda \mathcal{L}_{\text{ortho}}
\end{equation}
where $\lambda$ is a hyperparameter controlling the strength of this regularization term.
\begin{table}[!t]
\caption{The experiment results of AKT-Rec and baselines. Best results are in bold.}
\label{tab:base_results}
\resizebox{\columnwidth}{!}{%
\begin{tabular}{ccccccc}
\hline
                        & \multicolumn{2}{c}{Total}              & \multicolumn{2}{c}{Head}          & \multicolumn{2}{c}{Tail} \\ \cline{2-7} 
\multirow{-2}{*}{Model} & AUC    & GAUC   & AUC    & GAUC        & AUC         & GAUC       \\ \hline
Online base             & 0.7510 & 0.6385 & 0.7528 & 0.6477      & 0.7485      & 0.6137     \\ \hline
SaviorRec               & 0.7521 & 0.6455 & 0.7534 & 0.6516      & 0.7507      & 0.6347     \\
TailNet                 & 0.7491 & 0.6370 & 0.7509 & 0.6453      & 0.7479      & 0.6448           \\
POSO                    & 0.7518 & 0.6412 & 0.7520 & 0.6472      & 0.7497      & 0.6321     \\
SimTier                 & 0.7515 & 0.6398 & 0.7529 & 0.6481      & 0.7496      & 0.6279     \\ \hline
AKT-Rec & \textbf{0.7536} & \textbf{0.6483} & \textbf{0.7543} & \textbf{0.6528} & \textbf{0.7522} & \textbf{0.6397} \\ \hline
\end{tabular}%
}
\end{table}
\section{Experiments}
\subsection{Experimental Setup}
\subsubsection{Dataset and Metrics}
We evaluate the proposed framework on an industrial dataset collected from the Tmall mobile application. The dataset contains two months of click logs from June 2025 to August 2025, covering 36 million users and about 300 million items. We use the last five days as the test set and the remaining days as the training set. We define long-tail users as users with fewer than five interactions in the training set, which accounts for 85.58\% of users. We define long-tail items as items exposed fewer than ten times in the training set, which accounts for 95.8\% of items. We define long-tail samples as samples where either the user or the target item is a long-tail ID, which accounts for 22.4\% of samples. Following prior work, we report Area Under the ROC Curve (AUC) and Group AUC (GAUC) for offline evaluation. For online A/B testing, we report Clicks, CTR, Click-Through Conversion Rate (CTCVR), and Gross Merchandise Value (GMV).

\subsubsection{Implementation Details}
We build a co-occurrence database from co-occurrence signals on the Tmall platform and compute item features from the training set for MLLM prompts. We use GME-Qwen2-VL-7B~\cite{zhang2024gme} to encode item multi-modal content into semantic representations. For user semantic representations, we fine-tune Qwen3-30B-A3B~\cite{qwen3technicalreport} with supervision and extract user representations. For user/items semantic IDs, we use RQ-VAE with three codebooks of size 128/256.

\subsubsection{Baselines}
We compare AKT-Rec with strong baselines, including multimodal cold-start methods such as SaviorRec~\cite{yao2025saviorrec} and SimTier~\cite{sheng2024enhancing}, and non-multimodal methods like POSO~\cite{dai2021poso}, TailNet~\cite{liu2020long}, and our online base model.
 
\subsection{Overall Performance}
As shown in Table~\ref{tab:base_results}, AKT-Rec achieves consistently higher AUC and GAUC than all baselines across different user activity levels. The gains are most pronounced on long-tail samples. Compared with the online base model, AKT-Rec improves AUC by 0.346\% and GAUC improves by 1.53\%. On head samples, AKT-Rec still maintains an advantage, which indicates that the method improves long-tail modeling without sacrificing accuracy on head IDs.
\subsection{Ablation Study}
\begin{table}[]
\caption{Ablation study of AKT-Rec}
\label{tab:ablation}
\resizebox{\columnwidth}{!}{%
\begin{tabular}{@{}ccccccc@{}}
\toprule
\multirow{2}{*}{Model}       & \multicolumn{2}{c}{Full} & \multicolumn{2}{c}{Head} & \multicolumn{2}{c}{Tail} \\ \cmidrule(l){2-7} 
                    & AUC     & GAUC    & AUC     & GAUC    & AUC     & GAUC    \\ \midrule
w/o individual emb. & -0.46\% & -0.92\% & -0.67\% & -0.93\% & -0.42\% & -0.44\% \\
w/o cluster emb.    & -0.19\% & -0.82\% & -0.03\% & -0.17\% & -0.44\% & -1.20\% \\
w/o CGAE gate       & -0.13\% & -0.50\% & -0.26\% & -0.22\% & -0.07\% & -0.12\% \\ 
w/o   instance-level feature & -1.14\%     & -1.55\%    & -1.38\%     & -1.81\%    & -0.83\%     & -1.21\%    \\
w/o cluster-level feature    & -0.17\%     & -0.34\%    & -0.12\%     & -0.21\%    & -0.26\%     & -0.63\%    \\
w/o HFA gate        & -0.13\% & -0.30\% & -0.31\% & -0.55\% & -0.11\% & -0.23\% \\
\bottomrule
\end{tabular}%
}
\end{table}
We conducted comprehensive ablation studies to assess their respective contributions as detailed in Table~\ref{tab:ablation}.
Removing the individual embedding and representing user and item IDs only with the shared cluster embedding reduces overall AUC by 
0.46\% and GAUC by 
0.92\%, which confirms the necessity of the individual component in AKT-Rec. Conversely, removing the cluster embedding yields a clear performance drop, with a larger degradation on tail samples, suggesting that tail IDs benefit disproportionately from cluster-level knowledge. Replacing the activity-based gate in CGAE with a simple average further decreases AUC by 0.13\% and GAUC by 0.5\%. Removing instance-level features triggers a sharp decline in AUC (1.14\%) and GAUC (1.55\%), particularly for head samples, which confirms the necessity of individual behavioral sequences. In contrast, omitting cluster-level features leads to a 0.17\% AUC and 0.34\% GAUC drop, with tail samples exhibiting higher sensitivity to this information. Finally, replacing the HFA activity gate with a fixed average ($\alpha = 0.5$) degrades performance, demonstrating that adaptive fusion weights are critical to maximize information utility across both head and tail IDs.
\subsection{Online A/B Testing}
We conduct a two-week online A/B test on the Tmall platform, where both the experimental and control groups are allocated 10\% of the online traffic. Table~\ref{tab:online_result} summarizes the performance gains. Specifically, AKT-Rec yields improvements of 2.73\% in clicks, 2.76\% in CTR, 1.7\% in CTCVR, and 3.47\% in GMV. These results demonstrate the effectiveness of AKT-Rec in addressing long-tail distribution challenges within industrial recommendation scenarios.
\begin{table}[!h]
\caption{The online A/B test results of AKT-Rec}
\label{tab:online_result}
{%
\begin{tabular}{c|cccc}
\hline
Online Metrics & Clicks  & CTR     & CTCVR  & GMV     \\ \hline
Gain(\%)       & +2.73\% & +2.76\% & +1.7\% & +3.47\% \\ \hline
\end{tabular}%
}
\end{table}
\section{Conclusion}
This paper presents AKT-Rec, a long-tail recommendation framework built on generative semantic IDs. AKT-Rec uses LLMs to extract semantic representations for items and users, and applies a RQ-VAE to discretize these representations into semantic IDs, which naturally form semantic clusters. Within the feature fusion module, Cluster-Guided Adaptive Embedding and Hierarchical Feature Aggregation enable asymmetric knowledge transfer from head to tail IDs. Extensive offline and online experiments on large-scale datasets demonstrate that AKT-Rec consistently outperforms all baselines and significantly enhances recommendation performance in real-world industrial scenarios.

\bibliographystyle{ACM-Reference-Format}
\bibliography{references}

\section*{Presenter Bio}
\textbf{Chenyi Yan}  is an Algorithm Engineer at Alibaba Group, where he develops core recommendation algorithms for the homepage of the Tmall App. His current research focus lies at the intersection of large language models and recommender systems, specifically overcoming long-tail recommendation challenges in large-scale e-commerce scenarios. Chenyi Yan is dedicated to translating cutting-edge AI research into scalable, real-world solutions that significantly enhance personalized user experiences and drive business growth.

\end{document}